\documentclass[lettersize,journal]{IEEEtran}
\usepackage{amsmath,amsfonts}
\usepackage{algorithmic}
\usepackage{algorithm}
\usepackage{array}
\usepackage[font=normalsize,labelfont=sf,textfont=sf]{subcaption}
\usepackage{textcomp}
\usepackage{stfloats}
\usepackage{verbatim}
\usepackage{graphicx}
\usepackage{cite}
\usepackage{xcolor}
\usepackage{makecell}
\usepackage[utf8]{inputenc} % allow utf-8 input
\usepackage[T1]{fontenc}    % use 8-bit T1 fonts    
\usepackage{url}            % simple URL typesetting
\usepackage{booktabs}       % professional-quality tables
\usepackage{amsfonts}       % blackboard math symbols
\usepackage{nicefrac}       % compact symbols for 1/2, etc.
\usepackage{microtype}      % microtypography
\usepackage{xcolor}         % colors
\usepackage{multirow}
\usepackage{tikz}
\usetikzlibrary{svg.path}      % <-- needed for svg{...} paths
\usepackage{scalerel}  
\usepackage{scalerel}
\usepackage{float}
\usepackage[subtle]{savetrees} 
\usepackage{hyperref}   % hyperlinks

\definecolor{orcidlogocol}{HTML}{A6CE39}
\tikzset{
  orcidlogo/.pic={
    \fill[orcidlogocol]
      svg{M256,128c0,70.7-57.3,128-128,128C57.3,256,0,198.7,0,128C0,57.3,57.3,0,128,0C198.7,0,256,57.3,256,128z};
    \fill[white]
      svg{M86.3,186.2H70.9V79.1h15.4v48.4V186.2z}
      svg{M108.9,79.1h41.6c39.6,0,57,28.3,57,53.6c0,27.5-21.5,53.6-56.8,53.6h-41.8V79.1z M124.3,172.4h24.5c34.9,0,42.9-26.5,42.9-39.7c0-21.5-13.7-39.7-43.7-39.7h-23.7V172.4z}
      svg{M88.7,56.8c0,5.5-4.5,10.1-10.1,10.1c-5.6,0-10.1-4.6-10.1-10.1c0-5.6,4.5-10.1,10.1-10.1C84.2,46.7,88.7,51.3,88.7,56.8z};
  }
}

\newcommand\orcidicon[1]{\href{https://orcid.org/#1}{\mbox{\scalerel*{
                \begin{tikzpicture}[yscale=-1,transform shape]
                \pic{orcidlogo};
                \end{tikzpicture}
            }{|}}}}

\begin{document}

\title{HRTFformer: A Spatially-Aware Transformer for Individual HRTF Upsampling in Immersive Audio Rendering}

\author{Xuyi~Hu\textsuperscript{$\ast$~\orcidicon{0009-0005-8938-8527}},~\IEEEmembership{Member,~IEEE,} 
Jian~Li\textsuperscript{$\ast$~\orcidicon{0009-0003-2274-3489}}, 
Shaojie~Zhang\,\textsuperscript{\orcidicon{0009-0005-9007-3661}},  
Stefan~Goetz\,\textsuperscript{\orcidicon{0000-0002-1944-0714}},~\IEEEmembership{Member,~IEEE,} Lorenzo~Picinali\,\textsuperscript{\orcidicon{0000-0001-9297-2613}}, \\ Ozgur~B.~Akan\,\textsuperscript{\orcidicon{0000-0003-2523-3858}},~\IEEEmembership{Fellow,~IEEE,} and~Aidan~O.~T.~Hogg\,\textsuperscript{\orcidicon{0000-0001-5501-7799}}

\thanks{This study was made possible by support from SONICOM (www.sonicom.eu), a project that has received funding from the European Union’s Horizon 2020 research and innovation program under grant agreement No. 101017743.}
\thanks{$^\ast$ These authors contributed equally to this work.}

}
        % <-this % stops a space
%\thanks{Manuscript received April 19, 2021; revised August 16, 2021.}}

% \author{
% Xuyi~Hu\,\textsuperscript{\orcidicon{0009-0005-8938-8527}}, %~\IEEEmembership{Member,~IEEE,}
%         Jian~Li, 
%          ~Lorenzo~Picinali\,\textsuperscript{\orcidicon{0000-0001-9297-2613}},
%         and~Aidan~O.~T.~Hogg\,\textsuperscript{\orcidicon{0000-0001-5501-7799}}}
% \thanks{
% This study was made possible by support from SONICOM (www.sonicom.eu), a project that has received funding from the European Union’s Horizon 2020 research and innovation program under grant agreement No. 101017743.}

\markboth{Journal of \LaTeX\ Class Files,~Vol.~14, No.~8, August~2021}%
{Shell \MakeLowercase{\textit{et al.}}: A Sample Article Using IEEEtran.cls for IEEE Journals}

\maketitle
\IEEEpubid{\begin{minipage}{\textwidth}\ \\\\\\\\[12pt] \centering
  This work has been submitted to the IEEE for possible publication. \\Copyright may be transferred without notice, after which this version may no longer be accessible.
\end{minipage}}

\begin{abstract}
Individual Head-Related Transfer Functions (HRTFs) are starting to be introduced in many commercial immersive audio applications and are crucial for realistic spatial audio rendering. However, one of the main hesitations regarding their introduction is that creating individual HRTFs is impractical at scale due to the complexities of the HRTF measurement process. To mitigate this drawback, HRTF spatial upsampling has been proposed with the aim of reducing the measurements required. While prior work has seen success with different machine learning (ML) approaches, these models often struggle with long-range preservation of local spatial variation patterns across neighbouring source directions and generalization at high upsampling factors. In this paper, we propose a novel transformer-based architecture for HRTF upsampling, leveraging the attention mechanism to better capture spatial correlations across the HRTF sphere. Working in the spherical harmonic (SH) domain, our model learns to reconstruct high-resolution HRTFs from sparse input measurements with significantly improved accuracy. To enhance spatial coherence, we introduce a neighbour dissimilarity loss that promotes magnitude smoothness, yielding more realistic upsampling. We evaluate our method using both perceptual localization models and objective spectral distortion metrics. Experiments show that our model outperforms existing methods across several evaluation metrics in generating realistic, high-fidelity HRTFs.  
\end{abstract}

\begin{IEEEkeywords}
immersive audio, head-related transfer function, transformer, upsampling, interpolation
\end{IEEEkeywords}

\section{Introduction}
Immersive audio often plays a vital role in applications such as virtual reality (VR)~\cite{herre2023mpeg, quackenbush2021mpeg}, augmented reality (AR)~\cite{steadman2019short}, gaming~\cite{sun2021immersive,brahimaj2024enhancing}, and even therapeutic contexts~\cite{silva2016perceiving} where it aims to recreate realistic spatial soundscapes that align with human auditory perception. 
Human spatial hearing relies on interaural and monaural localization cues. Interaural cues are typically categorized as interaural time differences (ITDs), which dominate at low frequencies, and interaural level differences (ILDs), which dominate at mid–high frequencies~\cite{picinali2022system}. In the regions often termed as the `cone of confusion'~\cite{moore2012introduction}, where different source locations yield similar ITDs and ILDs, the auditory system exploits monaural spectral cues shaped by the pinnae~\cite{ick2024spatially}. The centre frequency, depth, and placement of these pinna-induced spectral notches provide crucial information for elevation and for resolving front–back ambiguity. As expected, these spectral cues, together with ITDs and ILDs that depend on the listener’s head-and-torso morphology, are highly unique to each listener. These cues can all be captured by a person's Head-Related Transfer Function (HRTF), which describes how an individual's anatomy filters sound from different directions before it reaches the eardrums~\cite{bruschi2024,mouchtaris2002inverse,jin2013creating}. 
% A critical component to achieve this realistic spatial sound rendering is that of a Head-Related Transfer Function (HRTF). This is because HRTFs capture how an individual’s anatomy, particularly the head, torso, and pinnae, filters incoming sound from different directions before it reaches the eardrums~\cite{bruschi2024}.
% These effects occur due to the interaction of sound waves with the listener’s head, torso, and pinnae, including reflections and diffractions that take place before the sound arrives at the ear canal. HRTFs encapsulate these transformations and encode both binaural cues~\cite{gray2021transmission} and monaural spectral cues that are critical for accurate sound localization.

It is well known that using non-individualized HRTFs, which are not personally tailored to a listener, can significantly compromise spatial audio performance. Perceptual studies have shown that mismatched HRTFs can impair sound-source localization, particularly in the sagittal plane, where listener-specific spectral cues are important.

They can also reduce perceptual qualities such as externalization, realism, and immersion, and may negatively affect task performance in virtual environments. These findings motivate the need for HRTF personalization in immersive audio systems. In terms of HRTF personalization, various methods have been proposed showing that generic HRTFs can often lead to impaired sound source localization, as the accurate spectral cues needed for spatial perception are strongly influenced by individual anatomical features, particularly the shape of the listeners' pinnae~\cite{stitt2019auditory, lu2025bicg}. In addition to causing localization errors, non-individualized HRTFs have also been shown to negatively impact perceptual qualities such as externalization~\cite{baumgartner2021decision}, immersion, timbral coloration, realism, and spatial depth~\cite{majdak2014acoustic,armstrong2018perceptual}. Furthermore, the use of poorly matched HRTFs can reduce a listener’s ability to segregate and understand speech in complex auditory scenes, including multi-talker environments or in the cocktail party scenario~\cite{oehler2023importance,poirier2018impact}. These drawbacks highlight the need and importance of personalization of HRTFs to be able to deliver accurate and immersive auditory experiences 
~\cite{hogg2021polynomial,guezenoc2020hrtf,zhang2020individual,marggraf2025impact}.

\begin{figure*}[t] \centering
    \includegraphics[width=\textwidth]{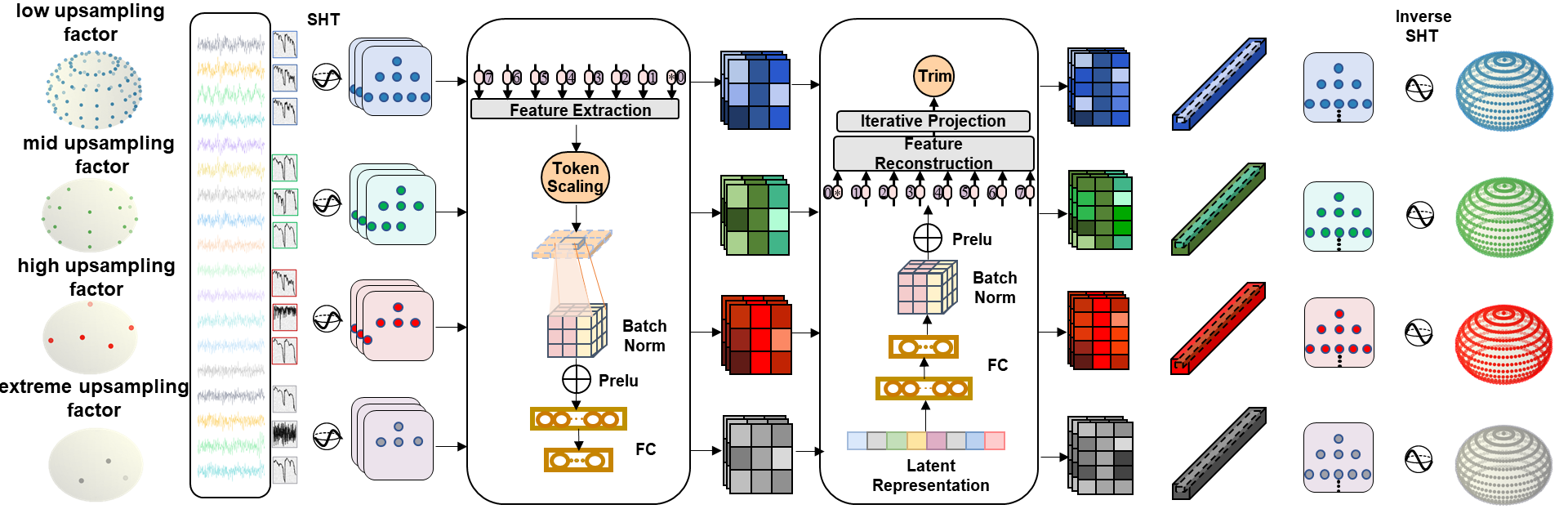}
    \caption{The HRTF upsampling workflow of HRTFformer: the low-resolution HRTFs are first transformed into the spherical harmonics domain, which provides a compact and physically meaningful representation of directional acoustic information. The resulting SH coefficients serve as the model's input. The encoder extracts global and local spatial features from the input and compresses them into the latent representation, which the decoder then uses to extrapolate and interpolate higher-degree SH coefficients. These are finally converted back into high-resolution HRTFs. This process is applied under four different sparsity conditions to evaluate the model's robustness. \label{fig:Framework}}
    \vspace{-0.3cm}
\end{figure*}

Various approaches to HRTF personalization have been proposed, including direct acoustic measurements 
~\cite{engel2023sonicom}, 3D surface scanning 
~\cite{dinakaran2018perceptually,zhang2020modeling,pirard2025enhancing,pirard2026photogrammetry}, anthropometry-based models~\cite{grijalva2016manifold, wang2020global,zhang2023grid}, and selection from databases of measured HRTFs~\cite{marggraf2024hrtf}. Among these, taking a direct acoustic measurement still remains the `gold standard', as it is able to capture the listener-specific filtering effects precisely using in-ear microphones and controlled speaker arrays~\cite{braren2021towards,lee2022hrtf,niwa2018efficient}. However, this approach is time-consuming, requires specialized equipment, and must be conducted in a noise-free environment, limiting its scalability and motivating the development of more practical alternatives.

To alleviate some of the downsides and difficulties with taking a direct acoustic measurement of a spatially dense HRTF, HRTF spatial upsampling has emerged as a promising alternative approach~\cite{siripornpitak2022spatial,zhao2025head}. It aims to reconstruct high-resolution HRTFs from a sparse set of acoustic measurements, significantly reducing the number of required sampling points. By leveraging spatial correlations and the underlying structure in HRTF data, upsampling methods enable efficient personalization while reducing measurement time and hardware needs.

HRTF spatial upsampling techniques are commonly divided into interpolation and learning-based approaches. Interpolation-based approaches estimate HRTFs at new source positions by combining existing measurements, for example through weighted sums of measured impulse responses or by representing the HRTF field using continuous spatial basis functions such as spherical harmonics (SH)~\cite{bau2022estimation,zhang2021individualized}.
However, their performance degrades significantly when only sparse measurements are available, as they rely heavily on dense, evenly distributed input data. These approaches are also based on fixed mathematical assumptions, such as smoothness or symmetry, which may not accurately capture the complex, subject-specific variations in real HRTFs. These limitations have motivated the development of machine learning (ML) approaches - including supervised learning using morphological features~\cite{fernandez2023prediction,mccarthy2025machine}, neural networks trained on spatial HRTF data~\cite{masuyama2025retrieval,ko2023prtfnet,lu2025deep}, perception-aware personalization in the spherical harmonic domain~\cite{zheng2026perceptual}, and generative models that synthesize individual HRTFs from sparse measurements~\cite{hogg2024hrtf,xia2023upmix,niu2025head}. Such approaches demonstrate better generalization and reconstruction quality, especially under sparse sampling conditions. However, they typically require large amounts of high-resolution training data and may struggle to generalize to unseen listeners. On the matter of data availability, at the time of writing, there are approximately 15 HRTF datasets openly available online~\cite{sofa_freefield_databases}, each comprising between 30 and 300 separate individuals. There is, however, significant evidence that data from different datasets are clearly distinguishable~\cite{pauwels2023relevance,daugintis2025listener}, due to the nature of the measurement system and procedure, and therefore attention should be paid when using multiple datasets for training.

% learning-based methods. In contrast, learning-based approaches synthesize HRTFs for novel source positions using neural networks trained on high-resolution datasets~\cite{hogg2023exploring,miccini2020hrtf}. Such approaches demonstrate better generalization and reconstruction quality, especially under sparse sampling conditions. However, they typically require large amounts of high-resolution training data, which can be difficult to obtain. Learning-based approaches also struggle to generalize to new listeners if the training set lacks anatomical diversity, and there's also a risk of overfitting to specific subjects. 

% Machine learning (ML) approaches have recently emerged as a promising direction for advancing HRTF personalization. Existing ML-based techniques include supervised learning using morphological features~\cite{fernandez2023prediction}, neural networks trained on spatial HRTF data~\cite{masuyama2025retrieval,ko2023prtfnet}, and generative models that synthesize personalized HRTFs from sparse measurements~\cite{hogg2024hrtf,xia2023upmix}. By leveraging data-driven representations, ML methods offer promising solutions for scalable, fast, and individualized HRTF estimation without requiring extensive physical measurements.

This paper introduces HRTFformer (shown in Fig.~\ref{fig:Framework}), a spatially-aware transformer designed for HRTF upsampling. HRTFformer aims to overcome one of the main challenges that current ML methods face, which is that although they often achieve low log-spectral distortion (LSD), this frequently does not correlate well with perceptual performance (which is usually overlooked). It was found in~\cite{hogg2025listener} that the most state-of-the-art methods under high sparsity tend to generate general (average) HRTFs, which contradicts the goal of individual customization. To address this issue, our HRTFformer approach adopts an attention mechanism in transformers to model dependencies across spherical harmonics and therefore enable the model to better capture global spatial correlations of the sound energy distribution, which may improve perceptual performance.Furthermore, we introduce a neighbour dissimilarity loss to enforce spatial continuity in the magnitude spectrum. Together, these components enable more perceptually coherent HRTF upsampling results. It is shown in this work that HRTFformer is able to create more realistic and individual HRTFs when compared to the latest state-of-the-art methods, especially in extremely sparse scenarios. 

In summary, the contributions of this paper are as follows:
\begin{enumerate}
    \item We propose a transformer-based architecture tailored for HRTF upsampling, named HRTFformer, which effectively captures global dependencies between sound energy distribution patterns, overcoming the limitations of performance degradation with spatially sparse data.
    \item We introduce a novel neighbour dissimilarity loss to enhance preservation of local spatial variation patterns across neighbouring source directions by respecting the natural variation in adjacent directions, thereby improving the realism and personalization of the reconstructed HRTFs.
    \item We evaluate HRTFformer on both sparse and dense measurements, showing robust reconstruction performance.
    \item We conduct comprehensive evaluations and demonstrate that HRTFformer achieves state-of-the-art results in both spectral evaluation and perceptual localization accuracy.
\end{enumerate}

\section{Related Work}
It is common to separate existing methods into two categories: those that are Interpolation-based and those that are Learning-based or data-driven.
\subsection{Interpolation-Based Approaches}
Among interpolation-based methods, barycentric interpolation~\cite{li2023barycentric,berrut2004barycentric} and SH interpolation~\cite{suda2002fast,hu2025machine} are commonly used. Barycentric interpolation estimates missing HRTFs by computing a weighted average of the three nearest neighbours, performing well when measurements are densely sampled (e.g., every 10–15°). However, its accuracy declines with sparser inputs (e.g., 30–40° spacing) due to increased distance between reference points. 
Similarly, SH interpolation represents the HRTF as a weighted sum of spatially continuous basis functions, the SHs. The SH coefficients (i.e. the contribution of each SH) are estimated by fitting this expansion to the measurements on the HRTF sphere (typically, least squares, sometimes weighted). When the number of samples is small relative to the chosen SH order, the fit becomes ill-conditioned: high-order terms can overfit noise and measurement error, leading to spurious spectral notches. In practice, therefore, the SH order must be limited based on the amount of sparse data available, and regularization or physics-based priors are added. However, SH interpolation will always struggle to capture higher frequency spectral content when the data is sparse. 
% Similarly, SH interpolation depends on global averaging assumptions and struggles to capture fine spatial details when data is sparse. 
% These methods depend on averaging known data points using predefined assumptions. For example, barycentric interpolation estimates values by weighting the three nearest neighbors, but its accuracy deteriorates as the distance between these neighbors increases.

\subsection{Learning-based Approaches}
Recent advancements in ML have opened up promising avenues for HRTF personalization~\cite{zhang2023hrtf,ito2025spatial,chen2026exploring,hu2026graph}. These data-driven methods aim to model the complex relationship between an individual's anatomical features and their corresponding HRTFs~\cite{masuyama2024niirf,gebru2021implicit,lee2023global,ma2023spatial}. Methods based on autoencoder architectures~\cite{ito2022head,jiang2023modeling,hu2024hrtf,chen2025spatial,hu2025hrtf} emphasize the frequency-domain characteristics of HRTFs by encoding them into compact latent representations. In addition to latent-space modeling, spherical CNN-based HRTF interpolation methods have been explored to better exploit the inherent geometry of HRTFs distributed over the sphere~\cite{chen2023head}. However, the upsampling performance has shown limited improvement over the interpolation methods. Generative Adversarial Network (GAN) based models have demonstrated strong capabilities in reconstructing missing information from sparsely sampled HRTFs~\cite{hogg2024hrtf,zhao2025head,chen2026spatial}. By learning complex spatial and spectral patterns from a rich set of high-resolution HRTFs during training, these models can effectively infer plausible high-resolution outputs, even when the input measurements are limited. At least 4–5 measurements are required due to architectural constraints of the model, which limit its applicability in extremely sparse conditions. Moreover, LSD results often fail to align with perceptual evaluations for learning-based methods. In several cases, models that achieved strong performance in terms of LSD exhibited notably poor outcomes in perceptual assessments, highlighting a disconnect between objective metrics and subjective audio quality~\cite{geronazzo2024technical,fantini2025survey}.

\subsection{Transformer Models}
% Transformers introduced self-attention to model long-range dependencies without recurrence, instead using positional encodings to represent sequence order. This led to transformers rapidly becoming popular for natural language processing, along with becoming the state-of-the-art for many speech and audio applications~\cite{zaman2025transformers,dong2018speech,gulati2020conformer}. The audio framework consists typically of a lightweight front end that converts waveforms or spectrograms into embeddings. The transformer can then use these embeddings to capture global context across spatial directions and spectral components. This flexibility motivates their use for HRTF upsampling, especially where \textcolor{red}{preservation of local spatial variation patterns across neighbouring source directions is essential.}

Transformers introduced self-attention to model long-range dependencies without recurrence, instead using positional encodings to represent sequence order. This led to transformers rapidly becoming popular for natural language processing, along with becoming the state-of-the-art for many speech and audio applications~\cite{zaman2025transformers,dong2018speech,gulati2020conformer}. The audio framework consists typically of a lightweight front end that converts waveforms or spectrograms into embeddings. The transformer can then use these embeddings to capture global context across spatial directions and spectral components. Beyond conventional sequence modeling, recent works have also adapted attention and convolutional architectures to spherical domains. SphereFormer~\cite{lai2023spherical} introduces radial-window self-attention for LiDAR point clouds, improving information aggregation under non-uniform spherical sampling. Attention mechanisms defined directly on the sphere have also been proposed to incorporate geometric priors such as spherical topology and rotational structure~\cite{bonevattention}. The flexibility of attention mechanisms therefore motivates their use in transformer-based HRTF upsampling models, particularly where both long-range dependencies across SH coefficients and local spatial variation patterns across neighbouring source directions need to be preserved.

\begin{figure*}[t] \centering
    \includegraphics[width=\textwidth]{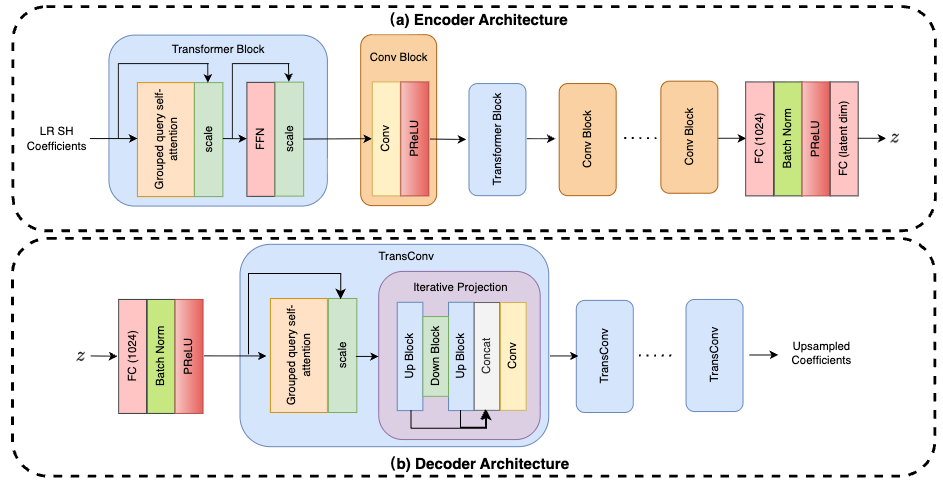}
    \caption{The model architecture of our HRTFformer. The encoder integrates transformer layers with convolutional downsampling modules to progressively extract and compress spatial features from low-resolution SH coefficients into the latent representation. The decoder combines transformer layers with iterative projection units that perform upsampling.} \label{fig:model architecture}
    \vspace{-0.3cm}
\end{figure*}

\section{Method}
\subsection{Data Pre-processing}
In general, HRTF data points are sampled on the surface of a sphere, making them inherently three-dimensional. To accommodate transformer architectures, which usually require one-dimensional sequential inputs, spherical harmonics transformation (SHT)~\cite{mohlenkamp1999fast} is utilized to project the HRTF data $f(\theta, \phi)$ onto a series of orthogonal spherical harmonic basis functions and corresponding coefficients. The resulting SH coefficients $F^m_l$ of degree $l$ and order $m$ are computed as:
\vspace{-0.1cm}
\begin{equation}
F_{l}^{m} = \int_{0}^{2\pi} \int_{0}^{\pi} f(\theta, \phi) Y_{l}^{m}(\theta, \phi) \sin(\phi) d\phi d\theta\;, 
\end{equation}
where $\theta$ and $\phi$ represent the azimuth and elevation angles, respectively. In acoustic applications, the SH basis function is defined as:
\begin{equation}
Y_{l}^{m}(\theta, \phi) = \sqrt{\frac{(2l+1)(l-m)!}{4\pi(l+m)!}} P_{l}^{m}(\cos(\phi)) e^{jm\theta}\;, 
\end{equation}
where $P_{l}^{m}(x)$ are the associated Legendre functions. The inverse SHT reconstructs the original HRTF function from its SH coefficients $F_{l}^{m}$ via the following expression:
\vspace{-0.1cm}
\begin{equation}
f(\theta, \phi) = \sum_{l=0}^{\infty} \sum_{m=-l}^{l} F_{l}^{m} Y_{l}^{m}(\theta, \phi)\;. 
\end{equation}
This SH representation offers multiple advantages: the coefficients naturally form a sequential structure compatible with transformer-based models. Each basis function represents a unique pattern of sound energy distribution in the space, and such a physically meaningful decomposition aligns well with the highly directional nature of HRTFs and facilitates more effective modeling and upsampling.

\subsection{Model Structure}\label{sec:model structure}
As illustrated in  Fig.~\ref{fig:model architecture}, the proposed model adopts an encoder-decoder framework. The encoder takes low-resolution SH coefficients as input, extracts hierarchical features, and compresses them into a latent representation $\textbf{z}$ of dimensionality 128. Subsequently, the decoder maps $\textbf{z}$ back to high-resolution SH coefficients.

The structure of the encoder is depicted in Fig.~\ref{fig:model architecture}(a). It consists of alternating transformer blocks and convolutional blocks, where transformer blocks capture global contextual relationships between coefficients via self-attention mechanisms, while convolutional blocks perform local feature extraction and feature map downsampling. Grouped query self-attention enables each head to focus on distinct spatial patterns, enhancing feature extraction in the transformer blocks.
% Grouped query self-attention is employed in transformer blocks, which is critical in feature extraction since separate attention heads can effectively focus on distinct frequency components or spatial sound energy distribution patterns.
% while reducing the memory and computational costs compared to ordinary multi-head attention by sharing key-value heads across grouped queries.

Position encoding is vital in our task since low-degree spherical harmonics are typically more important than high-degree ones. However, since attention computation is permutation-invariant, explicit positional cues must be included. Therefore, rotary position embedding (RoPE)~\cite{su2024roformer} is adopted. RoPE incorporates relative positional information by rotating query and key vectors, allowing the model to capture both local and global positional relationships.  Compared to absolute position encodings~\cite{vaswani2017attention, dosovitskiy2020image}, RoPE better generalizes to variable input resolutions and provides a more natural way to model the ordering of SH coefficients. Consider a $d$-dimensional vector $x$ at position index $p$, the rotary position embedding is applied by multiplying by a rotation matrix:
\begin{align}
\text{RoPE}(x, p) &=
\begin{bmatrix}
\cos \theta_{i,p} & -\sin \theta_{i,p} \\
\sin \theta_{i,p} & \cos \theta_{i,p}
\end{bmatrix}
\begin{bmatrix}
x_{2i} \\
x_{2i+1}
\end{bmatrix}, \nonumber \\
&\quad \text{for } i = 0, 1, \dots, \tfrac{d}{2}-1
\end{align}

\noindent where $\theta_{i,p}$ is the rotation angle.  

\begin{equation}
Q' = \text{RoPE}(Q, p), 
\quad 
K' = \text{RoPE}(K, p),
\end{equation}

\vspace{-0.2cm}
\begin{equation}
\text{Attention}(Q', K', V) 
= \text{softmax}\!\left( \frac{Q'K'^{\top}}{\sqrt{d_k}} \right)V.
\end{equation}

% the relative position encoding~\cite{shaw2018self, liu2021swin, jiang2021transgan} is adopted, which better handles variable input resolutions and naturally balances the global and local positional information more effectively than absolute encodings~\cite{vaswani2017attention, dosovitskiy2020image}. The relative position information is incorporated by adding a learnable bias term $E$ to the attention score:
% \begin{equation}
%     Attention(Q,K,V) = softmax(((\frac{QK^T}{\sqrt{d_k}} + E) V)
% \end{equation}
% where $Q,K,V\in \mathbb{R}^{S \times C}$ represent query, key, value matrices, $S,C$ denote the sequence length and channels of the feature map. The bias term $E$ is taken from a learnable relative position table $M \in \mathbb{R}^{2L-1}$ where $L$ corresponds to the maximum number of coefficients.

While layer normalization is commonly used in transformer architectures~\cite{vaswani2017attention, liu2021swin, zhang2021multi} for training stability, we empirically observe that the token scaling approach proposed in~\cite{jiang2021transgan} demonstrates superior performance in our task (please refer to Table \ref{tab:ablation}). Unlike layer normalization, token scaling preserves the relative energy distribution across frequency bins by avoiding mean subtraction, which is especially crucial in SH coefficient modeling because these coefficients are magnitude-sensitive and their relative scale carries physically and perceptually meaningful information. Token scaling can be written as:
\begin{equation}
    y = \frac{x}{\sqrt{\frac{1}{C} \sum^{C-1}_{i=0}(x^i)^2 + \epsilon}}.
\end{equation}
where $x$ denotes the input feature, $y$ represents the scaled output, $C$ indicates the number of frequency channels, and $\epsilon$ is a small constant for numerical stability.

% The architecture of the decoder is illustrated in Fig.~\ref{fig:model architecture}(b). The decoder, as shown in  Fig.~\ref{fig:model architecture}(b), maintains a similar structure with the encoder, but replaces convolutional blocks with iterative projection units for upsampling. This design adopts the iterative resolution refinement strategy proposed by~\cite{haris2018deep}, where repeated up- and downsampling operations enable robust learning of the mapping between low- and high-resolution representations of SH coefficients. The arrows indicate residual connections. Transformer layers are also employed throughout the decoder to capture complex, long-range dependencies across different SH degrees and frequency components, which significantly improves interpolation and extrapolation performance, leading to finer high-resolution HRTF reconstruction results.

The architecture of the decoder is illustrated in Fig.~\ref{fig:model architecture}(b). To address the upsampling challenge, we replace the standard feedforward layer in the transformer block with an iterative projection mechanism. This design adopts the iterative resolution refinement strategy proposed by~\cite{haris2018deep}, where repeated up- and downsampling operations enable robust learning of the mapping between low- and high-resolution representations of SH coefficients. The arrows indicate residual connections, which further stabilize the training. Meanwhile, the attention module captures complex, long-range dependencies across different SH degrees and frequency components, which significantly improves interpolation and extrapolation performance, leading to finer high-resolution HRTF reconstruction results.

\subsection{Loss Functions}
The overall loss function contains three terms: LSD, ILD, and neighbour dissimilarity loss (NDL). Therefore, the complete loss function can be shown as:
\vspace{-0.14cm}
\begin{equation}
    \mathcal{L} = \text{LSD} + \text{ILD} + \text{NDL}, 
\end{equation}

The LSD measures the discrepancy in magnitude between the reconstructed HRTF $H_{\text{G}}$ and ground-truth high-resolution HRTF $H_{\text{HR}}$. It is defined as:
\vspace{-0.1cm}
\begin{equation}
    \text{LSD} = \frac{1}{N}\sum^N_{n=1}\sqrt{\frac{1}{W}\sum^W_{w=1}\left( 20\text{log}_{10} \frac{|  H_{\text{HR}}(f_w, x_n)   |}{ | H_{\text{G}}(f_w, x_n) | } \right)^2} \;,   
    \label{eq:LSD}
\end{equation}
where $N$ represents the total number of spatial positions, and $W$ denotes the total number of frequency bins. $H_{\text{HR}}$ and $H_{\text{G}}$ are the targeted high-resolution HRTF and generated HRTF respectively.

The ILD quantifies the difference in magnitude between the left and right ear responses of an HRTF set. The ILD loss evaluates the deviation between the ILD values of the reconstructed and reference HRTFs, expressed as: 
\vspace{-0.2cm}
\begin{align}
\text{ILD} &= \frac{1}{N}\sum_{n=1}^N \frac{1}{W}\sum_{w=1}^W 
    \Bigg| 20\log_{10}\!\left(\frac{H^{\text{Left}}_{\text{HR}}(f_w, x_n)}
                                     {H^{\text{Right}}_{\text{HR}}(f_w, x_n)}\right) \nonumber \\
&\quad - 20\log_{10}\!\left(\frac{H^{\text{Left}}_{\text{G}}(f_w, x_n)}
                                   {H^{\text{Right}}_{\text{G}}(f_w, x_n)}\right) 
    \Bigg| \;.
\end{align}

The NDL is employed to encourage smooth spatial variation in the magnitude of the generated HRTF. For each spatial position, the neighbour dissimilarity measures the deviation between the HRTF at that position and the average of its four immediate neighbours. Assuming spatial continuity of HRTFs, this deviation should remain small. The neighbour dissimilarity loss quantifies the discrepancy between the neighbour dissimilarity patterns of the generated HRTF and the target HRTF, calculated as:

\vspace{-0.2cm}
\begin{align}
\mathcal{L}_{\text{ND}} &= \frac{1}{N}\sum_{n=1}^N
\Bigg(
\Bigg( H_{HR}^{(n)} - \frac{1}{|\mathcal{K}(n)|}\sum_{k\in\mathcal{K}(n)} H_{HR}^{(k)} \Bigg) \nonumber \\
&\quad -
\Bigg( H_G^{(n)} - \frac{1}{|\mathcal{K}(n)|}\sum_{k\in\mathcal{K}(n)} H_G^{(k)} \Bigg)
\Bigg)^2,
\end{align}where $\mathcal{K}(n)$ represents the set of connected neighbourhoods of position $n$, $|\mathcal{K}(n)|$ is the number of neighbouring points (four in our case), and $N$ is the total number of spatial positions.

Although the HRTF measurement positions lie on a spherical surface, the SONICOM sampling grid uses azimuth angles that are consistently defined across elevation angles. This allows the sampled positions to be indexed approximately as a regular 2D grid for the purpose of computing the NDL. Accordingly, the four neighbours of each point are defined as the adjacent samples in elevation and azimuth, i.e., the top, bottom, left, and right positions. This is an approximation to the true spherical geometry, but it provides a simple and consistent way to encourage local spatial smoothness.

% \vspace{-0.2cm}
% \begin{align}
% \mathcal{L}_{\text{ND}} &= \frac{1}{N}\sum_{n=1}^N
% \Bigg( 
% \Bigg( H_{HR}^{(n)} - \frac{1}{|\mathcal{K}(n)|}\sum_{k\in\mathcal{K}(n)} H_{HR}^{(k)} \Bigg) \nonumber \\
% &\quad -
% \Bigg( H_G^{(n)} - \frac{1}{|\mathcal{K}(n)|}\sum_{k\in\mathcal{K}(n)} H_G^{(k)} \Bigg)
% \Bigg)^2 .
% \end{align}

% where $\mathcal{K}(n)$ represents the set of connected neighborhoods of position n, $|\mathcal{K}(n)|$  is the number of neighboring points, which is 4 in our case. N is the total number of spatial positions.

\begin{table*}[t]
\centering
    \caption{ITD, ILD, LSD evaluation results for sparsity level 3, 5, 19, and 100. Best-performing results are in bold.}
\label{tab:spatial cue results}

\resizebox{1.\textwidth}{!}{
    \begin{tabular}{lcccccccccccc}

    \toprule
    %{} & \multicolumn{6}{c}{Dataset A} & \multicolumn{4}{c}{Dataset B} \\
    %\cmidrule[0.5pt](rl){2-7}
    %\cmidrule[0.5pt](rl){8-11}
    {} & \multicolumn{3}{c}{Sparsity level 3}  & \multicolumn{3}{c}{Sparsity level 5} & \multicolumn{3}{c}{Sparsity level 19}  & \multicolumn{3}{c}{Sparsity level 100} \\
    \cmidrule[0.5pt](rl){2-4}
    \cmidrule[0.5pt](rl){5-7}
    \cmidrule[0.5pt](rl){8-10}
    \cmidrule[0.5pt](rl){11-13}
        {Method} & {ITD} & {ILD} & {LSD} & {ITD} & {ILD} & {LSD} & {ITD} & {ILD} & {LSD} & {ITD} & {ILD} & {LSD}  \\
    \midrule
    GEP-GAN~\cite{hogg2024hrtf} &  36.64 & 1.14 & 5.20 & 33.40 & 1.15 & 4.41 & 37.25 & 1.35 & 4.10 & 33.41 & 0.48 & 3.20 \\
    IOA3D~\cite{zhao2025head} &  22.84 & 1.00 & 4.67 & 16.00 & 0.75 & 4.90 & 13.95 & 0.69 & 3.21 & \textbf{6.96} & 0.41 & \textbf{2.10} \\
    SYT-FSP-AE~\cite{ito2022head} &  24.66 & 1.28 & 4.42 & 18.32 & 1.07 & 4.36 & 21.38 & 0.91 & 3.25 & 17.83 & 0.76 & 2.21 \\
    Kalimotxo~\cite{arevalo2025spatial} &  30.46 & 0.92 & 4.49 & 31.39 & 0.72 & 4.85 & 25.08 & 0.81 & 3.29 & 21.18 & 0.73 & 3.06 \\
    AE-GAN~\cite{hu2024hrtf} &  32.38 & 1.20 & 4.79 & 27.66 & 1.18 & 4.57 & 22.19 & 1.41 & 3.45 & 27.76 & 0.66 & 2.58 \\
    SH~\cite{arend2023magnitude} &  78.61 & 6.05 & 9.96 & 77.14 & 5.44 & 10.35 & 62.75 & 1.68 & 5.43 & 47.67 & 0.44 & 3.38 \\
    Barycentric~\cite{cuevas20193d} &  49.05 & 7.50 & 8.56 & 47.86 & 4.54 & 8.33 & 45.37 & 1.76 & 4.79 & 41.32 & 0.55 & 3.20 \\
    HRTFformer (Ours) &  \textbf{17.50} & \textbf{0.75} & \textbf{4.20} & \textbf{15.29} & \textbf{0.64} & \textbf{4.18} & \textbf{13.45} & \textbf{0.67} & \textbf{3.10} & 19.36 & \textbf{0.38} & 3.15 \\
    \bottomrule
    \end{tabular}
}
\end{table*}

\begin{figure*}[t]
\centering

\makebox[0.22\textwidth]{\footnotesize  Sparsity level 3}
\makebox[0.22\textwidth]{\footnotesize  Sparsity level 5}
\makebox[0.22\textwidth]{\footnotesize  Sparsity level 19}
\makebox[0.22\textwidth]{\footnotesize  Sparsity level 100}

%\makebox[\textwidth]{\small Sparsity level 3} \\
% \includegraphics[width=1\textwidth,height=2.5cm]{SYT-FSP-AE_LSD.pdf} 

% \includegraphics[width=1\textwidth,height=2.5cm]{Kalimotxo_LSD.pdf} 

% \includegraphics[width=1\textwidth,height=2.5cm]{IOA3D_LSD.pdf} 

% \includegraphics[width=1\textwidth,height=3cm]{HRTFformer_LSD.pdf} 
\includegraphics[width=1\textwidth,height=2.6cm]{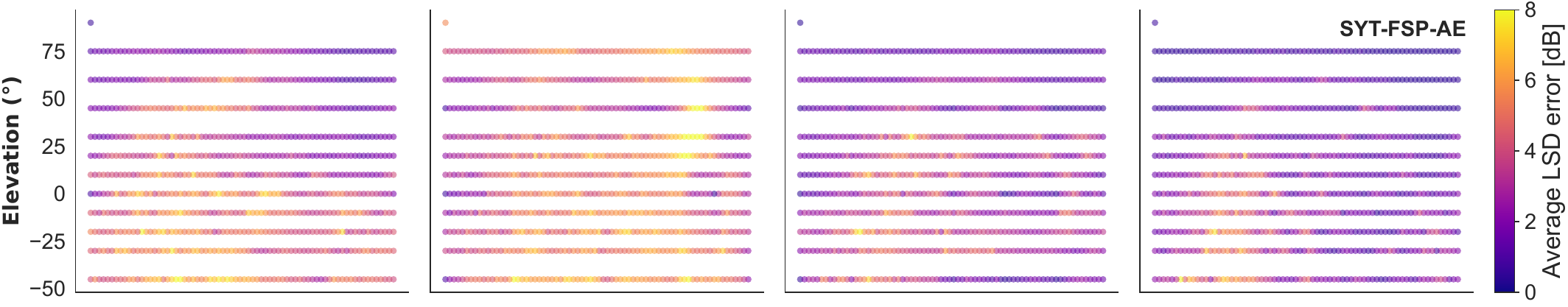}
\includegraphics[width=1\textwidth,height=2.6cm]{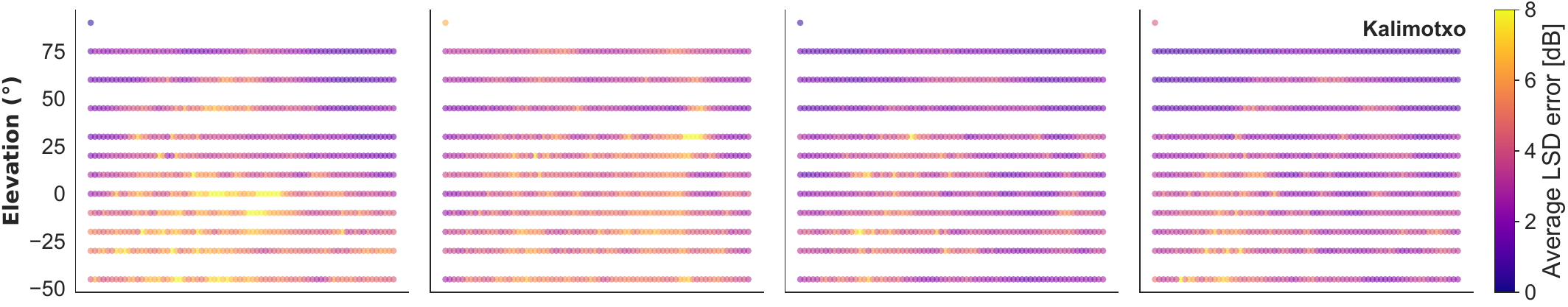}
\includegraphics[width=1\textwidth,height=2.6cm]{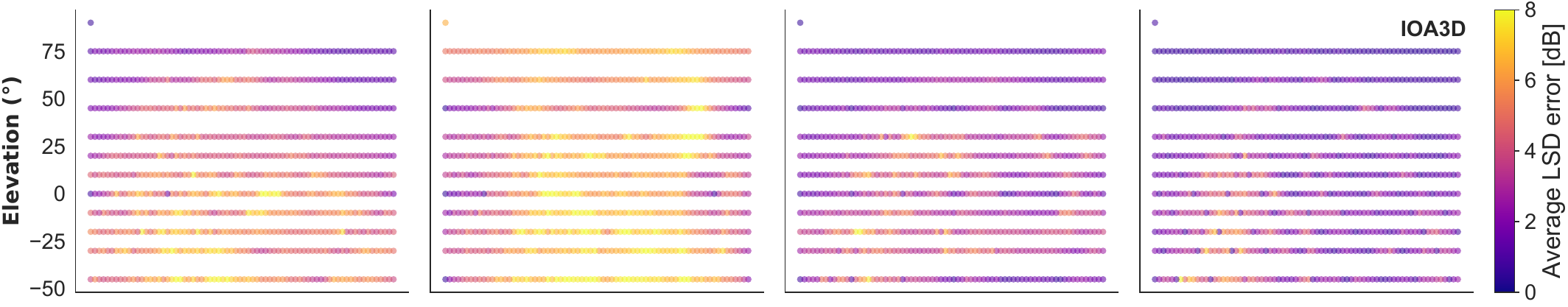}
\includegraphics[width=1\textwidth,height=3cm]{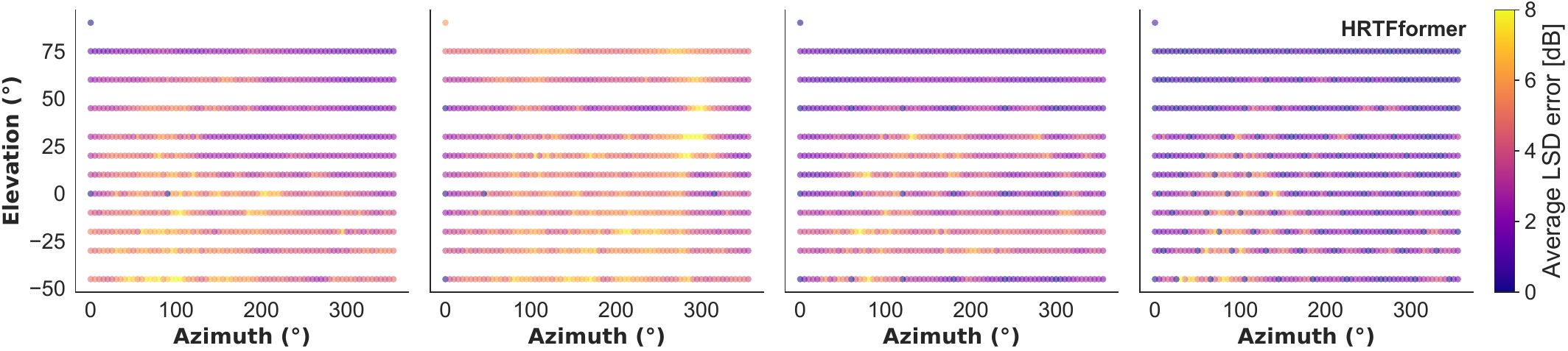} 
\caption{LSD distributions for selected subject across various HRTF upsampling methods and sparsity levels. Top to bottom: SYT-FSP-AE, Kalimotxo, IOA3D, and HRTFformer.}

\vspace{-0.3cm}
\label{fig:LSD plot}
\end{figure*}

\section{Experiments}
\subsection{Implementation Details}
In the experiments, the strides settings within the Conv Block were adjusted to accommodate varying input sizes according to different sparsity levels. The model was trained using a batch size of 8, and a learning rate of 0.0002 for 200 epochs and optimized using the Adam optimizer. All training was conducted on a single NVIDIA RTX 4090 GPU with 24GB of memory.

\subsection{Tasks and Dataset}
% We evaluated our model on The SONICOM HRTF dataset~\cite{engel2023sonicom} under four sparsity levels, using 3, 5, 19, and 100 initial sampling points. To comprehensively assess the performance, We employed both spatial cue metrics, including LSD, ILD, and interaural time difference (ITD), as well as perceptual localization metrics. Out model was compared with nine baselines, including traditional non-machine learning methods such as barycentric and spherical harmonics interpolation, and deep-learning approaches such as GEP-GAN and Kalimotxo.

We evaluated our model on the SONICOM HRTF dataset~\cite{engel2023sonicom,poole2025extended} under four sparsity levels—using 3, 5, 19, and 100 initial sampling points. To comprehensively assess performance, we employed several spatial cue metrics, including LSD, ILD, and ITD, which reflect the accuracy of binaural localisation cues, as well as perceptual localization metrics, which estimate how well listeners can localise sound sources based on the reconstructed HRTFs. Our model was compared against nine baselines, including interpolation (non-ML) methods such as barycentric interpolation, SH and SUpDEq~\cite{arend2019spatial} methods, and deep learning models such as AE-GAN~\cite{hu2024hrtf} and Kalimotxo~\cite{arevalo2025spatial}.

\subsection{Result of Spatial Cue Evaluation}
As shown in Table \ref{tab:spatial cue results}, our HRTFformer consistently achieves the lowest errors across all three spatial cue metrics (ITD, ILD, and LSD) under the most challenging sparsity levels (3 and 5), significantly outperforming both learning-based and traditional interpolation baselines. This indicates that our model is robust in modeling spatial acoustic patterns even under extremely limited sampling conditions. Furthermore, at low sparsity levels (sparsity levels 100), HRTFformer obtains comparable results, where it performs on par with existing state-of-the-art methods. Although HRTFformer excels at high sparsity levels, its advantage diminishes with denser inputs, likely due to the transformer's emphasis on global patterns over local details.
% While HRTFformer shows clear advantages at high sparsity levels, the relative improvement narrows as the number of initial points increases. This limitation may stem from the nature of the attention mechanism in the transformer architectures, which primarily focuses on global correlations between directional sound distribution patterns and may overlook local details. 

To better understand which locations these errors come from, Fig.~\ref{fig:LSD plot} is a visualization of the spectral discrepancy across elevation and azimuth angles for four selected methods under four sparsity levels. Although all existing methods can capture HRTF spatial information and reconstruct it well at a low sparsity level, high sparsity levels represent more practical scenarios, as they significantly reduce the time and effort required from users during HRTF acquisition. As the sparsity level increases (moving leftward), baseline approaches show a noticeable rise in reconstruction errors, especially at sparsity level 3. In contrast, our model maintains significantly lower error levels, demonstrating its superior generalization in sparse settings due to its transformer-based architecture.

To quantitatively separate errors introduced by the SH representation from those arising from the model itself, we first take the 41 original high-resolution HRTF measurements in the test set, represented in the HRTF domain. These HRTF measurements are then transformed into the SH domain using the same SH order adopted by the model. The resulting SH representation is subsequently reconstructed back into the HRTF domain via the inverse SH transform. Finally, the ITD, ILD, and LSD are computed between the original high-resolution HRTF and the SH-encoded-then-decoded HRTF, as reported in Table~\ref{tab:Intrinsic_SH_error}. The results show that the SH representation explains only a minor part of the ITD and LSD errors (30$\%$ and 26$\%$, respectively). Its contribution to the ILD error is even smaller, accounting for only 10$\%$ on average. Since the model is trained with a loss computed in the HRTF domain, the predicted SH coefficients are optimized after inverse SH reconstruction to minimize the discrepancy between the predicted and target HRTFs. As a result, the model can partially compensate for errors arising from sparse-input reconstruction and SH-domain prediction. However, it cannot recover information that is intrinsically lost due to the finite target SH order or incomplete spatial sampling. Therefore, when the full-pipeline error approaches the SH-only error, this indicates that the additional error introduced by the model is small and that the remaining discrepancy is mainly due to the inherent limitations of the SH representation.

\begin{table}[t]
\centering
\caption{Intrinsic SH error as a percentage of the total error.}
\label{tab:Intrinsic_SH_error}
\setlength{\tabcolsep}{2pt}
\footnotesize
\resizebox{0.48\textwidth}{!}{%
\begin{tabular}{cccc}
    \toprule
    \makecell{SONICOM\\ID} &
    \makecell{ITD error\\explained by SH (\%)} &
    \makecell{ILD error\\explained by SH (\%)} &
    \makecell{LSD error\\explained by SH (\%)} \\
    \midrule
    P068  & 23 & 14 & 20 \\
    P082  & 24 & 6 & 22 \\
    P0104 & 32 & 14 & 25 \\
    \makecell{Average (41 samples)} & 30 & 10 & 26 \\
    \bottomrule
\end{tabular}%
}
\end{table}

% At sparsity level 100, where all models perform well with LSD below 4dB, most regions appear dark purple, showing accurate reconstruction. Only a small cluster of higher errors is visible between azimuth $20^\circ$ and $120^\circ$, corresponding to the subject's right-front quadrant. As the upsampling factor increases (moving leftward), yellow regions become more pronounced, showing a general degradation in reconstruction quality due to reduced sampling. At sparsity levels 5 and 19, high errors occur evenly in the space. However, at sparsity level 3, the error distribution varies between methods. For HRTFformer, the yellow regions are substantially less compared to other methods, especially in the higher elevation angles ($>50\circ$) and azimuth angles $>200\circ$. This demonstrates the robustness of our model in reconstructing HRTFs even in extremely sparse conditions. On the other hand, models such as SYT-FSP-AE and Kalimotxo exhibit considerably higher errors in the frontal regions, particularly at azimuths below $50\circ$ and above $300\circ$, highlighting their reduced adaptability in sparse settings. Notably, a similar error distribution to ours is observed in IOA3D, suggesting that both models may capture comparable directional features, yet HRTFformer demonstrates superior generalization due to its transformer-based architecture.

\begin{figure*}[t] \centering
    \includegraphics[width=\textwidth]{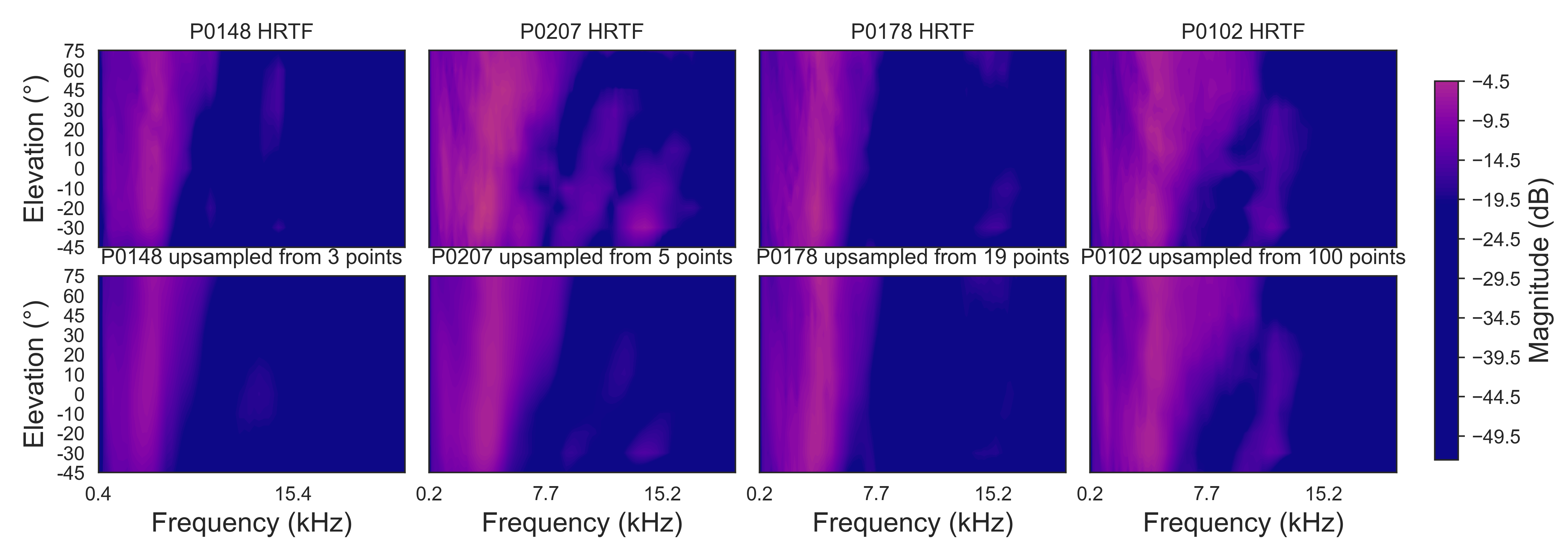}
    \caption{Median plane spectra of example upsampled HRTFs using HRTFformer compared to the original HRTFs for each upsampling factor.} \label{fig:Median plane spectra}
    \vspace{-0.3cm}
\end{figure*}

\begin{table*}[t]
\centering
    \caption{Perceptual evaluation results in sparsity level 3 and 5. Best-performing results are in bold.}
\label{tab:perceptual1}

\resizebox{1.\textwidth}{!}{
    \begin{tabular}{lcccccc}

    \toprule
    %{} & \multicolumn{6}{c}{Dataset A} & \multicolumn{4}{c}{Dataset B} \\
    %\cmidrule[0.5pt](rl){2-7}
    %\cmidrule[0.5pt](rl){8-11}
    {} & \multicolumn{3}{c}{Sparsity level 3}  & \multicolumn{3}{c}{Sparsity level 5} \\
    \cmidrule[0.5pt](rl){2-4}
    \cmidrule[0.5pt](rl){5-7}
        {Method} & {Polar Accuracy Error} & {Polar RMS Error} & {Quadrant Error} & {Polar Accuracy Error} & {Polar RMS Error} & {Quadrant Error} \\
    \midrule
    GEP-GAN~\cite{hogg2024hrtf} &  5.66 & 47.60 & 16.35 & 5.22 & 46.06 & 28.17 \\
    IOA3D~\cite{zhao2025head} &  13.75 & 41.79 & 18.20 & 14.74 & 41.01 & 23.35 \\
    Kalimotxo~\cite{arevalo2025spatial} & 10.45 & 41.72 & 18.08 & 17.50 & 41.42 & 25.44 \\
    AE-GAN~\cite{hu2024hrtf} &  12.88 & 45.64 & 17.58 & 14.17 & 49.45 & 27.49 \\
    MERL2~\cite{masuyama2025retrieval} & 11.53 & 38.55 & 13.50 & 7.77 & 35.53 & 16.39 \\
    \toprule
    SH~\cite{arend2023magnitude} & 21.59 & 49.78 & 25.94 & 10.55 & 45.91 & 21.51 \\
    Barycentric~\cite{cuevas20193d} & 44.51 & 57.62 & 48.77 & 13.64 & 42.78 & 40.69 \\
    SUpDEq~\cite{arend2019spatial} & 16.53 & 47.15 & 27.96 & 21.15 & 38.36 & 20.22 \\
    \toprule
    HRTFformer (Ours) & \textbf{3.83} & \textbf{35.37} & \textbf{8.94} & \textbf{3.34} & \textbf{32.01} & \textbf{10.73} \\
    \bottomrule
    \end{tabular}
}
\end{table*}

\begin{table*}[t]
\centering
    \caption{Perceptual evaluation results in sparsity level 19 and 100. Best-performing results are in bold.}
\label{tab:perceptual2}

\resizebox{1.\textwidth}{!}{
    \begin{tabular}{lcccccc}

    \toprule
    %{} & \multicolumn{6}{c}{Dataset A} & \multicolumn{4}{c}{Dataset B} \\
    %\cmidrule[0.5pt](rl){2-7}
    %\cmidrule[0.5pt](rl){8-11}
    {} & \multicolumn{3}{c}{Sparsity level 19}  & \multicolumn{3}{c}{Sparsity level 100} \\
    \cmidrule[0.5pt](rl){2-4}
    \cmidrule[0.5pt](rl){5-7}
        {Method} & {Polar Accuracy Error} & {Polar RMS Error} & {Quadrant Error} & {Polar Accuracy Error} & {Polar RMS Error} & {Quadrant Error} \\
    \midrule
    GEP-GAN~\cite{hogg2024hrtf} & 8.06 & 45.68 & 24.89 & 20.00 & 43.87 & 22.14 \\
    IOA3D~\cite{zhao2025head} &  4.47 & 37.74 & 23.69 & 12.07 & 37.74 & 15.22 \\
    Kalimotxo~\cite{arevalo2025spatial} & 3.75 & 41.82 & 24.17 & 8.23 & 41.64 & 19.11 \\
    AE-GAN~\cite{hu2024hrtf} &  7.41 & 39.96 & 25.72 & 15.44 & 42.36 & 16.98 \\
    MERL2~\cite{masuyama2025retrieval} & 3.13 & 39.04 & 14.24 & \textbf{4.27} & 38.63 & \textbf{12.56} \\
    \toprule
    SH~\cite{arend2023magnitude} & 6.95 & 45.11 & 29.59 & 10.16 & 39.63 & 20.05 \\
    Barycentric~\cite{cuevas20193d} & 9.77 & 41.68 & 31.48 & 15.23 & 43.75 & 23.86 \\
    SUpDEq~\cite{arend2019spatial} & 1.60 & 41.03 & 15.70 & 10.89 & 36.85 & 15.70 \\
    \toprule
    HRTFformer (Ours) & \textbf{0.19} & \textbf{30.26} & \textbf{10.22} & 7.52 & \textbf{35.78} & 14.75 \\
    \bottomrule
    \end{tabular}
}
\end{table*}

\begin{figure*}[!tb]
    \centering

    % First row title
    \textbf{(a) Sparsity level 3}
    %\\[0.5em] % vertical space

    % First image (full-width)
    \includegraphics[width=\textwidth,height=3.5cm]{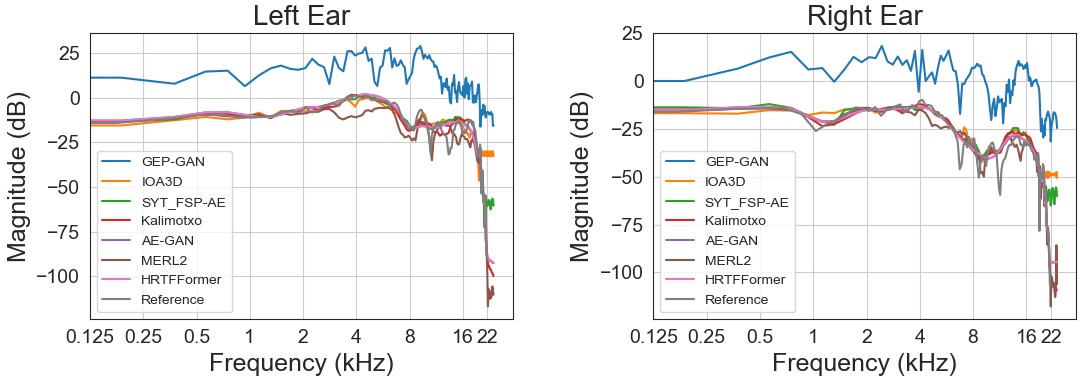}
    %\\[1em]  % space between images

    % Second row title
    \textbf{(b) Sparsity level 100}
    %\\[0.5em]

    % Second image (full-width)
    \includegraphics[width=\textwidth,height=3.5cm]{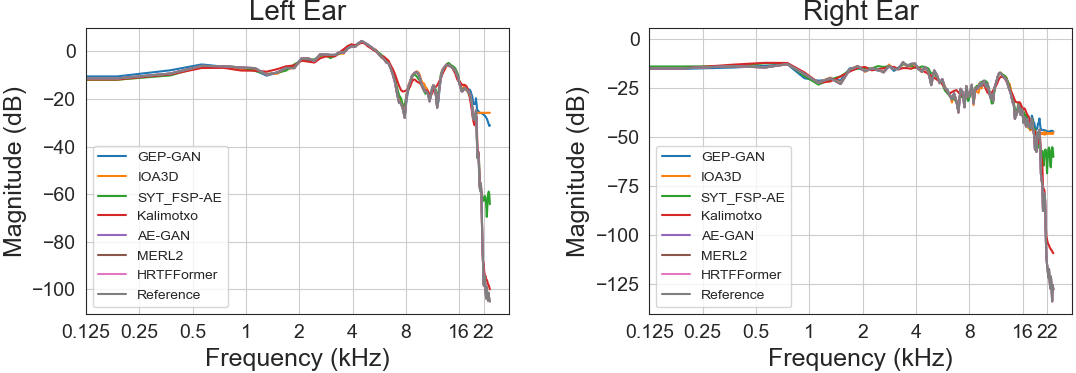}

    \caption{Upsampled HRTFs for subject P0203 with two sparsity levels, and a source to the right (45° azimuth, 0° elevation). The reference HRTF are shown for comparison.}
    \label{fig:Baselines_plot}
    \vspace{-0.3cm}
\end{figure*}

\subsection{Result of Perceptual localization Evaluation}
We employ the Bayesian auditory localization model~\cite{barumerli2023bayesian} proposed by Barumerli et al. to estimate localization performance from the reconstructed HRTFs for perceptual evaluation. This model predicts directional responses based on binaural and monaural auditory features and has been shown to be useful for evaluating the perceptual validity of HRTFs. Using this framework, we report Polar Accuracy Error, Polar RMS Error, and Quadrant Error. Polar Accuracy Error reflects the local bias in the polar dimension for responses within 90 degrees of the target, Polar RMS Error measures the overall magnitude of the local polar error for those same responses, and Quadrant Error gives the percentage of responses whose polar error exceeds 90 degrees, thereby capturing large top-down and front-back confusions. To avoid highly distorted polar errors at the far left and right sides of the listener, the polar and quadrant metrics are evaluated only for target directions within 30 degrees of the median plane.

% CITE: R. Barumerli, P. Majdak, M. Geronazzo, D. Meijer, F. Avanzini, and R. Baumgartner, “A Bayesian model for human directional localization of broadband static sound sources,” Acta Acustica, vol. 7, art. no. 12, 2023, doi: 10.1051/aacus/2023006.

Table \ref{tab:perceptual1} and \ref{tab:perceptual2} summarize the perceptual evaluation results. It should be noted that although Polar RMS Error values above 30 degrees may initially appear large, they are not necessarily unexpected from this model, because the metric is still evaluated over a broad range of median-plane directions, including front, elevated, and rear regions, where polar localisation is more challenging. This is also consistent with ~\cite{barumerli2023bayesian}, in which actual listener data show Polar Error values of 32.73 degrees. It is also consistent with \cite{hogg2024hrtf}, which uses the same evaluation framework, in which a Target condition is defined as the original high-resolution HRTF evaluated against itself (i.e. the best achievable performance), and this yields a Polar RMS Error of 32.11 degrees.

HRTFformer outperforms the baselines by a large margin in both polar accuracy error and quadrant error at sparsity levels 3, 5, and 19, demonstrating its effectiveness in preserving the high fidelity of reconstructed HRTFs, which is crucial for accurate sound localization. The comparable perceptual scores under sparsity level 100 suggest that spectral discrepancy does not always directly translate to perceptual differences. This observation is further supported by Fig.~\ref{fig:Baselines_plot}(a), where GEP-GAN, despite showing clear deviations in magnitude from the reference at sparsity level 3, still obtains competitive polar accuracy. At sparsity level 100, all methods manage to reconstruct spectral cues closely aligned with the reference for both ears, as shown in Fig.~\ref{fig:Baselines_plot}(b), reinforcing the notion that perceptual accuracy does not solely depend on spectral similarity. In addition, Fig.~\ref{fig:Median plane spectra} illustrates the spectral profiles of upsampled HRTFs generated by HRTFformer, compared against the ground truth HRTFs for each sparsity level at azimuth of $180^\circ$.

% \begin{figure}[tb]
%   \centering
%   \begin{subfigure}[b]{0.5\textwidth}
%     \centering
%     \includegraphics[height=0.3\textwidth,  width=1\textwidth]{ILD (1).png}
%     \caption{ILD for upsampled subject P0148 using HRTFFormer at sparsity level 3. Left: polar plot of absolute ILD across azimuth; Right: heatmap over azimuth-elevation locations.}
%     \label{fig:ild_sparsity3}
%   \end{subfigure}
%   % \hfill
%   \begin{subfigure}[b]{0.5\textwidth}
%     \centering
%     \includegraphics[height=0.3\textwidth, width=1\textwidth]{ITD (1).png}
%     \caption{ITD for upsampled subject P0148 using HRTFFormer at sparsity level 3. Left: polar plot of absolute ITD across azimuth; Right: heatmap over azimuth-elevation locations.}
%     \label{fig:itd_sparsity3}
%   \end{subfigure}
%   \caption{Comparison of interaural cues for upsampled subject P0148 using HRTFFormer at sparsity level 3. }
%   \label{fig:ITD&ILD}
% \end{figure}
\subsection{Computing Efficiency}
A comparison of runtime efficiency between different methods is reported in Table~\ref{tab:efficiency}. For third-party LAP challenge submissions without publicly available source code or full implementation details, computational metrics are reported as N/A to avoid unreliable estimates based on unsupported assumptions. HRTFformer has a larger number of parameters, model size, and inference time than the two GAN-based baselines~\cite{hu2025hrtf,hogg2024hrtf}. However, this increase in computational cost is not expected to be a limiting factor in the intended use case, since HRTFformer was not designed as a real-time system. Our primary objective was to maximize reconstruction quality and perceptual performance in HRTF upsampling, rather than to optimize computational efficiency. Moreover, HRTF upsampling is typically performed offline and therefore is not subject to strict real-time constraints. From this perspective, the additional complexity is compatible with the intended offline setting, particularly given the improved reconstruction and perceptual localization performance under sparse sampling conditions.

\begin{table}[t]
\centering
\caption{Comparison of computational efficiency in terms of the number of learnable weights, model size, FLOPs, and inference time.  N/A indicates values that cannot be reliably computed due to unavailable source code or implementation details.}
\label{tab:efficiency}
\setlength{\tabcolsep}{1.5pt} % reduce column padding
\small
\begin{tabular}{lcccc}
\toprule
Method & Param. (M) & Model (MB) & FLOPs (G) & Time (s) \\
\midrule
SYT-FSP-AE~\cite{ito2022head} & 5.68  & 21.7 & N/A & N/A \\
GEP-GAN~\cite{hogg2024hrtf}     & 137.62  & 525.10 & 91.36 & 0.05  \\
AE-GAN~\cite{hu2025hrtf}       & 179.95  &  576.94 & 105.72 & 0.09  \\
HRTFformer (Ours)& 270.60 & 1041.53 & 246.58 & 0.15  \\
\bottomrule
\end{tabular}
\end{table}

\subsection{Ablation Study}
We conducted a series of ablation experiments to evaluate the impact of different encoder and decoder structures, position embedding methods, normalization techniques, and loss functions (Table \ref{tab:ablation}).

\textbf{Model Structure}. As shown in Table \ref{tab:ablation}, incorporating transformer modules in both encoder and decoder consistently improves ITD, ILD, and LSD, showing the effectiveness of attention mechanisms in capturing dependencies of SH coefficients. In contrast, purely convolutional architectures primarily focus on local features, exhibit inferior performance. 
% In addition, perceptual metrics either improve or remain comparable, indicating that transformer-based architecture enhance spectral reconstruction quality without sacrificing perceptual localization accuracy.

\textbf{Position Embedding}. RoPE outperforms relative position bias~\cite{shaw2018self, liu2021swin, jiang2021transgan} across all metrics except for ILD. This suggests that encoding positional information directly into the query and key vectors is more effective than adding a learnable bias term to the attention scores.

\textbf{Normalization}. Token scaling achieves the best results, yielding the lowest LSD and ITD along with a substantial improvement in polar accuracy, proving the previous discussion in Sec. \ref{sec:model structure}.

\textbf{Loss Functions}. Ablation results on loss components reveal that combining LSD and ILD captures spectral distribution characteristics better than MSE. Adding NDL further enables the model to account for local magnitude variations and prevent abrupt changes, resulting in more realistic reconstruction, as evidenced by the lowest errors in both spatial cue and most perceptual metrics.

\begin{table*}[t]
\centering
\caption{Ablation study results in sparsity level 3. Best-performing results are in bold.}
\label{tab:ablation}

\resizebox{1.\textwidth}{!}{
    \begin{tabular}{llcccccc}
    \toprule
    \multirow{2}{*}{Component} & \multirow{2}{*}{Variations} & \multicolumn{3}{c}{Spatial Cue Evaluation} & \multicolumn{3}{c}{Perceptual Evaluation} \\
    \cmidrule(rl){3-5}
    \cmidrule(rl){6-8}
    & & {ITD} & {ILD} & {LSD} & {Polar Accuracy Error} & {Polar RMS Error} & {Quadrant Error} \\
    \midrule
    
    \multirow{2}{*}{Encoder} 
    & Resnet & 18.27 & 0.77 & 4.32 & 5.54 & 43.05 & \textbf{8.15} \\
    & Transformer & \textbf{17.50} & \textbf{0.75} & \textbf{4.20} & \textbf{3.83} & \textbf{42.37} & 8.94 \\
    % & GNN & 30.25 & 75.30 & 37.80 & 60.45 & 34.50 & 68.40 \\
    \midrule
    
    \multirow{2}{*}{Decoder} 
    & w/o Transformer & 18.10 & 1.43 & 5.54 & 5.67 & 42.53 & \textbf{6.81} \\
    & w/ Transformer & \textbf{17.50} & \textbf{0.75} & \textbf{4.20} & \textbf{3.83} & \textbf{42.37} & 8.94 \\
    \midrule
    
    \multirow{2}{*}{Position Embedding} 
    & Relative Position Bias & 18.25 & \textbf{0.66} & 4.23 & 6.08 & 42.82 & 8.94 \\
    & ROPE & \textbf{17.50} & 0.75 & \textbf{4.20} & \textbf{3.83} & \textbf{42.37} & \textbf{8.39} \\
    % & None & 29.40 & 74.20 & 37.20 & 59.80 & 33.90 & 67.50 \\
    \midrule
    
    \multirow{3}{*}{Normalization} 
    & LayerNorm & 17.74 & \textbf{0.66} & 4.25 & 5.65 & 43.05 & 9.81 \\
    & BatchNorm & 18.83 & 0.76 & 4.41 & 7.17 & 42.60 & \textbf{7.17} \\
    & Token Scaling & \textbf{17.50} & 0.75 & \textbf{4.20} & \textbf{3.83} & \textbf{42.37} & 8.94 \\
    \midrule
    
    \multirow{3}{*}{Loss Functions} 
    & MSE & 18.95 & 0.89 & 4.87 & 8.14 & 42.46 & \textbf{7.13} \\
    & LSD+ILD & 17.98 & 0.91 & 4.26 & 5.21 & 42.85 & 9.73 \\
    & LSD+ILD+NDL & \textbf{17.50} & \textbf{0.75} & \textbf{4.20}& \textbf{3.83} & \textbf{42.37} & 8.94 \\
    \bottomrule
    \end{tabular}
}
\end{table*}

\section{Conclusion and Future Work}
In this paper, we have proposed HRTFformer, a transformer-based model, to tackle the challenge of HRTF upsampling. It has been shown that by transforming HRTF data into the spherical harmonic domain and by leveraging attention mechanisms in transformer architecture, the model is able to learn the relationship between SH coefficients and, therefore, able to model the sound energy distribution pattern in the space effectively. A novel neighbour dissimilarity loss was also introduced to enforce spatial continuity in the HRTF magnitude spectrum across adjacent positions to achieve a more realistic HRTF reconstruction. The statistical results have suggested that HRTFformer is not only able to outperform other state-of-the-art methods in terms of objective metrics (LSD, ILD, and ITD) but also demonstrates improved perceptual evaluation results compared with the baselines, particularly under sparse sampling conditions, highlighting the model's effectiveness for use in real-life applications.

It is important to acknowledge a key limitation of the present work: all experiments were conducted using a single HRTF dataset, namely the SONICOM dataset. This choice was deliberate, motivated by recent challenges encountered when attempting to combine data from different HRTF repositories, which raised non-trivial consistency and compatibility issues~\cite{daugintis2025listener,pauwels2023relevance}. Nevertheless, future research will need to address this limitation more systematically. In particular, it will be essential to investigate strategies for enabling a degree of generalisation across datasets acquired under different measurement protocols and conditions. Such generalisability will be crucial for translating these methods into robust and scalable HRTF personalisation pipelines suitable for real-world applications.

In future work, we plan to confirm our perceptual results with subjective evaluations using real human listeners to assess spatial realism, externalization, and individual localization. This will overcome a current limitation of our results, which is that they rely on a perceptual model. It should also be noted that due to the nature of HRTF measurements being costly and time-consuming, datasets are limited, and, therefore, data diversity could have the potential to affect generalization to unseen subjects. To overcome this, we propose a future work that uses synthetic HRTFs for use in training via means of transfer learning, as this will likely help improve model generalization to unseen subjects and conditions.

% \section*{Acknowledgments}
% This should be a simple paragraph before the References to thank those individuals and institutions who have supported your work on this article.

% {\appendix[Proof of the Zonklar Equations]
% Use $\backslash${\tt{appendix}} if you have a single appendix:
% Do not use $\backslash${\tt{section}} anymore after $\backslash${\tt{appendix}}, only $\backslash${\tt{section*}}.
% If you have multiple appendixes use $\backslash${\tt{appendices}} then use $\backslash${\tt{section}} to start each appendix.
% You must declare a $\backslash${\tt{section}} before using any $\backslash${\tt{subsection}} or using $\backslash${\tt{label}} ($\backslash${\tt{appendices}} by itself
%  starts a section numbered zero.)}


\bibliographystyle{IEEEtran}
\bibliography{ref}

\newpage

\begin{IEEEbiography}[{\includegraphics[width=1in,height=1.25in,clip]{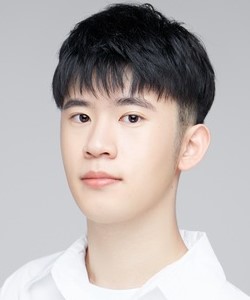}}]{Xuyi Hu} (Student Member, IEEE) is a PhD student at University of Cambridge. He studied Electrical and Electronic Engineering at University College London for his undergraduate degree and then completed a Master’s in Artificial Intelligence at Imperial College London. His research focuses on using machine learning to improve HRTF (head-related transfer function) upsampling and to measure personalised HRTFs. More information about current research projects can be found at: \url{https://georgehux.com/}.
\end{IEEEbiography}
% \vspace{-0.70cm}

\begin{IEEEbiography}[{\includegraphics[width=1in,height=1.25in,clip]{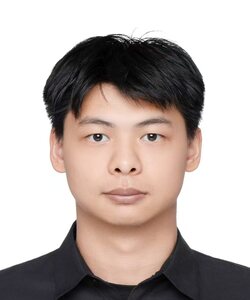}}]{Jian Li} is currently an employee at Huawei, specializing in the training of large language models. He holds a Bachelor’s degree in Electrical and Electronic Engineering from Nanyang Technological University (NTU) and a Master’s degree in Artificial Intelligence from Imperial College London.
\end{IEEEbiography}

\begin{IEEEbiography}[{\includegraphics[width=1in,height=1.25in,clip]{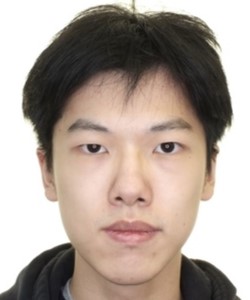}}]{Shaojie Zhang} (Member, IEEE) received the B.Eng. degree in electronics and electrical engineering from University College London in 2023, and the M.Phil. degree from the Department of Engineering, University of Cambridge in 2024, where he is currently pursuing the Ph.D. degree. His research interests include the applications of molecular communication systems and integrated sensing and communication within molecular communication.
\end{IEEEbiography}

\begin{IEEEbiography}[{\includegraphics[width=1in,height=1.25in,clip,keepaspectratio]{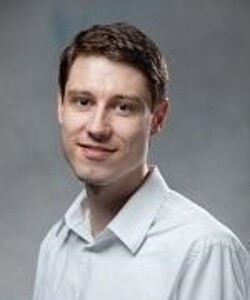}}]{Stefan Goetz}(Member, IEEE) studied physics and engineering. He received his undergraduate and graduate degrees from Technische Universitaet Muenchen (TU Muenchen), Munich, Germany, where he also conducted Ph.D. research on medical power electronics, partly at Columbia University, New York, NY, USA. He has worked in the automotive industry on electric drivetrains, machines, power electronics and vehicle architecture. He also contributed to the development of automotive chargers and grid integration technologies deployed worldwide. His research focuses on high-power electronics and magnetics for drive and medical applications, and on integrative power electronics for microgrids and e-vehicles.
\end{IEEEbiography}

\begin{IEEEbiography}[{\includegraphics[width=1in,height=1.25in,clip,keepaspectratio]{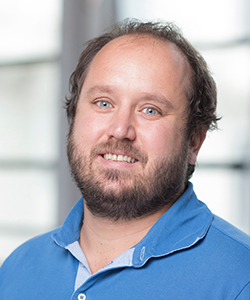}}]{Lorenzo Picinali} is a Professor in Spatial Acoustics and Immersive Audio and the lead of the Audio Experience Design at Imperial College London. In the past few years, he worked in Italy, France, and the United Kingdom on projects dealing with 3D binaural sound rendering, interactive applications for visually and hearing impaired individuals, audiology and hearing aid technology, audio and haptic interaction, and more in general, acoustical virtual and augmented reality. More information about the projects in which Lorenzo is involved can be found here: \url{https://www.axdesign.co.uk/.}
\end{IEEEbiography}

\begin{IEEEbiography}[{\includegraphics[width=1in,height=1.25in,clip,keepaspectratio]{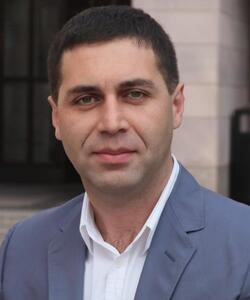}}]{Ozgur B. Akan}(Fellow, IEEE) received the Ph.D. degree from the School of Electrical and Computer Engineering, Georgia Institute of Technology, Atlanta, in 2004. He is currently the Head of the Internet of Everything Group, Department of Engineering, University of Cambridge, U.K., and the Director of the Centre for Next-Generation
Communications, Koç University, Türkiye. His research interests include wireless, nano, and molecular communications and Internet of Everything.
\end{IEEEbiography}

\begin{IEEEbiography}[{\includegraphics[width=1in,height=1.25in,clip,keepaspectratio]{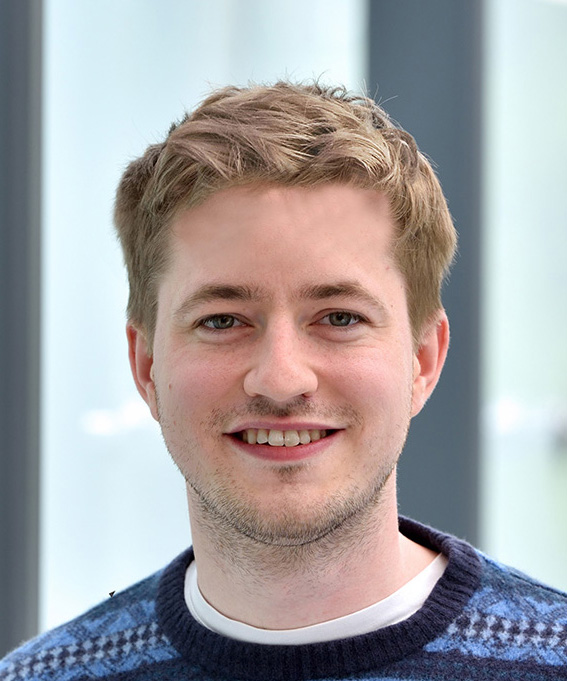}}]{Aidan O. T. Hogg} is a Lecturer in Computer Science and the co-lead for the Virtual, Immersive, Augmented and Binaural Audio Lab (VIABAL) in the Centre for Digital Music (C4DM) at Queen Mary University of London. He received an M.Eng. degree in electronic and information engineering and a PhD degree from Imperial College London in 2017 and 2022, respectively. His current research focuses on using deep learning to capture head-related transfer functions and, more generally, spatial acoustics and immersive audio. Other research interests include speaker diarization and statistical signal processing for audio applications. More information about current research projects can be found here: \url{https://aidanhogg.uk/}
\end{IEEEbiography}
% \vspace{-0.70cm}

\vfill

\end{document}